\documentclass[12pt,preprint]{aastex}

\shorttitle{The Subluminous Supernova 2007qd}
\shortauthors{McClelland et al.}

\begin{document}

\title{The Subluminous Supernova 2007qd: A Missing Link in a Family of
  Low-Luminosity Type Ia Supernovae}

\author{Colin M. M$^{\mathrm{c}}$Clelland\altaffilmark{1},
Peter M. Garnavich\altaffilmark{1},
Llu\'{i}s Galbany\altaffilmark{2},
Ramon Miquel\altaffilmark{2,3},
Ryan J. Foley\altaffilmark{4,5},
Alexei V. Filippenko\altaffilmark{6},
Bruce Bassett\altaffilmark{7},
J. Craig Wheeler\altaffilmark{8},
Ariel Goobar\altaffilmark{9,10},
Saurabh W. Jha\altaffilmark{11},
Masao Sako\altaffilmark{12},
Joshua A. Frieman\altaffilmark{13,14,15}, 
Jesper Sollerman\altaffilmark{16,17},
Jozsef Vinko\altaffilmark{18},
and
Donald P. Schneider\altaffilmark{19}
}
\altaffiltext{1}{Department of Physics, University of Notre Dame, Notre Dame, IN 46556.}
\altaffiltext{2}{Institut de F\'{i}sica d'Altes Energies, Universitat Aut\`{o}noma de Barcelona, E-08193 Barcelona, Spain.}
\altaffiltext{3}{Instituci\'o Catalana de Recerca i Estudis Avan\c{c}ats, E-08010 Barcelona, Spain.}
\altaffiltext{4}{Harvard-Smithsonian Center for Astrophysics, 60 Garden Street, Cambridge, MA 02138.}
\altaffiltext{5}{Clay Fellow.}
\altaffiltext{6}{Department of Astronomy, University of California, Berkeley, CA 94720-3411.}
\altaffiltext{7}{Department of Mathematics and Applied Mathematics, University of Cape Town, Rondebosch 7701, South Africa.}
\altaffiltext{8}{Department of Astronomy, McDonald Observatory, University of Texas, Austin, TX 78712.}
\altaffiltext{9}{Department of Physics, Stockholm University, AlbaNova University Center, SE-106 91 Stockholm, Sweden.}
\altaffiltext{10}{The Oskar Klein Centre for Cosmoparticle Physics, Department of Physics, AlbaNova, Stockholm University, SE-106 91 Stockholm, Sweden.}
\altaffiltext{11}{Department of Physics and Astronomy, Rutgers University, 136 Frelinghuysen Road, Piscataway, NJ 08854.}
\altaffiltext{12}{Department of Physics and Astronomy, University of Pennsylvania, 209 South 33rd Street , Philadelphia, PA 
19104.}
\altaffiltext{13}{Kavli Institute for Cosmological Physics, University of Chicago, 5640 South Ellise Avenue, Chicago, IL 60637.}
\altaffiltext{14}{Department of Astronomy and Astrophysics, University of Chicago, 5640 South Ellise Avenue, Chicago, IL 60637.}
\altaffiltext{15}{Center for Particle Astrophysics, Fermi National Accelerator Laboratory, P. O. Box 500, Batavia, IL 60510.}
\altaffiltext{16}{Dark Cosmology Centre, Niels Bohr Institute, University of Copenhagen, Juliane Maries Vej 30, DK-2100 Copenhagen, Denmark.}
\altaffiltext{17}{Department of Astronomy, The Oskar Klein Centre, Stockholm University, 10691 Stockholm, Sweden.}
\altaffiltext{18}{Department of Optics and Quantum Electronics, University of Szeged, Hungary.}
\altaffiltext{19}{Department of Astronomy and Astrophysics, 525 Davey Laboratory, Pennsylvania State University, University Park, PA 16802.}

\begin{abstract}

We present multi-band photometry and multi-epoch spectroscopy of the 
peculiar Type Ia supernova (SN~Ia) 2007qd, discovered by the SDSS-II 
Supernova Survey.  It possesses physical properties intermediate to 
those of the peculiar SN~2002cx and the extremely low-luminosity 
SN~2008ha.  Optical photometry indicates that it had an extraordinarily 
fast rise time of $\lesssim 10$~days and a peak absolute $B$ magnitude 
of $-15.4 \pm 0.2$ at most, making it one of the most subluminous 
SN~Ia ever observed.  Follow-up spectroscopy of SN~2007qd near 
maximum brightness unambiguously shows the presence of 
intermediate-mass elements which are likely caused by carbon/oxygen nuclear burning.  
Near maximum brightness, SN~2007qd had a photospheric velocity of 
only 2800~km~s$^{-1}$, similar to that of SN~2008ha but about 4000 
and 7000~km~s$^{-1}$ less than that of SN~2002cx and normal SN~Ia, 
respectively.  We show that the peak luminosities of SN~2002cx-like 
objects are highly correlated with both their light-curve stretch 
and photospheric velocities. Its strong apparent connection to 
other SN~2002cx-like events suggests that SN~2007qd is also a pure 
deflagration of a white dwarf, although other mechanisms cannot 
be ruled out.  It may be a critical link between SN~2008ha and 
the other members of the SN~2002cx-like class of objects.  
\end{abstract}


\section{Introduction}

For decades, Type Ia supernovae (SN~Ia) have been interpreted as
thermonuclear explosions that result from either accretion onto a
degenerate white dwarf or via coalescence of two degenerate stars; see
\citet{livio00} for a thorough review. Models that begin fusion
subsonically and then turn into a detonation are best at matching
observations of typical SN~Ia \citep{khokhlov91}.  Synthesized
intermediate-mass elements (IMEs) like silicon and sulfur in SN~Ia
spectra are the result of C and O burning at low densities due to the
pre-expansion of the white dwarf during the deflagration phase.  The
deflagration turns into a detonation, and the combination of the two
produces a significant amount of $^{56}$Ni, the decay of which powers
the light curve. 

SN~Ia show a range of energies and spectral characteristics, but the
majority of them are quite homogeneous when compared with other
supernova types.  The relative uniformity amongst SN~Ia explosions has
allowed their exploitation as cosmic standard candles, and observed
correlations between light-curve shape, peak luminosity
\citep{phillips93, hamuy96a}, and color \citep{riess96, guy07} have
enabled the precise measurement of cosmologically interesting
distances.  It is with these measurements, taken over a range of
redshifts, that estimates of the Hubble parameter ($H_0$) and cosmic
matter density have been refined \citep{hamuy95, riess95, garnavich98,
  freedman01, riess09}, and evidence for dark energy was revealed
\citep{riess98, perlmutter99}.

The Sloan Digital Sky Survey-II (SDSS-II) Supernova Survey was
designed to further improve SN~Ia as distance indicators
\citep{frieman08, sako08}.  Covering a 300 deg$^{2}$ area with a rapid
cadence, the survey discovered and spectroscopically confirmed roughly
500 SN~Ia over a range of redshifts out to $z \approx 0.4$.
\citet{kessler09a}, \citet{lampeitl09}, and \citet{sollerman09} used
the first year of SDSS-II data to constrain the dark energy
equation-of-state parameter $w = P/(\rho c^2)$, and analyzed the
systematic errors limiting the cosmological measurements.

Identifying the progenitors of SN~Ia and better understanding their
explosion mechanism may improve their reliability as distance
indicators. One approach is to study events that do not conform to the
general SN~Ia homogeneity in their spectra or luminosity. Subclasses
of SN~1991T-like \citep[hereafter referred to as 91T;][]{filippenko92b,
phillips92} and SN~1991bg-like \citep[hereafter
  91bg;][]{filippenko92a, leibundgut93, turatto96, garnavich04} events
were quickly recognized as being peculiar, although much of the
spectroscopic diversity is now known to be caused by a range of
photospheric temperatures \citep{nugent95}.  Recently, additional
subclasses of SN~Ia have been identified. For example, SN~2002ic
\citep{hamuy03} and SN~2005gj \citep{prieto05, aldering06} are
luminous objects that show hydrogen emission lines in their spectra,
unlike normal SN~Ia.  It is suspected they may be SN~Ia interacting
with dense circumstellar material, although a core-collapse scenario
has also been proposed \citep{benetti06}.

SN~2002cx (hereafter 02cx) was especially peculiar
\citep{filippenko03, li03}.  It showed a hot (91T-like) spectrum at
early times, but it cooled quickly after maximum brightness and its
expansion velocities were well below typical for a SN~Ia.  02cx
deviated significantly from the \citet{phillips93} relation between
peak luminosity and decline rate; it was subluminous for its
light-curve shape by $\sim 1.8$ mag in the $B$ and $V$ bands. The
well-observed SN~2005hk (hereafter 05hk) was spectroscopically very
similar to 02cx --- both exhibited photospheric velocities of roughly
7000~km~s$^{-1}$.  Photometrically, 05hk was faint on the Phillips
relation by $\sim 0.6$--1.0 mag in the optical, but $\sim 0.5$
mag more luminous than 02cx \citep{phillips07}.

\citet{jha06} have identified several other 02cx-like objects and
postulated that their extreme subluminosity suggests that they
constitute a class of pure thermonuclear deflagrations: that is, the
fusion front moving through the white dwarf fails to make the
transition to supersonic burning \citep{branch04} and is unable to
generate the large amounts of radioactive nickel observed in typical
SN~Ia \citep{foley09}.  The thermonuclear burning of carbon and oxygen
at moderate densities will create IMEs such as silicon, sulfur, and
calcium that dominate the spectrum.  Though typical SN~Ia experience
this phase only briefly, \citet{jha06} reasoned that 02cx-like objects
may burn completely via this mechanism. (For a list of 02cx-like
events, see Table~9 of \citealt{foley09}.)

Which characteristics separate delayed detonations from pure
deflagrations?  Simulations of pure deflagration events predict that
Rayleigh-Taylor instabilities will form clumps of radioactive nickel
and mix unburned material toward the center \citep{reinecke02,
  gamezo04, travaglio04, schmidt06}.  Late-time spectra of 02cx-like
events show many permitted iron transitions broken up into thin spikes
\citep{jha06}, indicative of the lower velocities that characterize
02cx-like supernovae.  These spectra also show narrow forbidden lines
of Ca~II and Fe~II, but are quite distinct from normal SN~Ia at late
epochs, whose spectra are dominated by relatively broad, blended Fe~II
and Fe~III lines.  Spectra of 05hk $\gtrsim$ 200 days past maximum
have been analyzed for emission from unburned C~I and O~I, but their
existence has not been confirmed \citep{sahu08}.

An extremely subluminous transient, SN~2008ha \citep[hereafter
  08ha;][]{foley09}, appears to be an additional member of the 02cx
class, although \citet{valenti09} have proposed that the extreme nature
of 08ha (and other 02cx-like objects) is better matched by the core
collapse of a massive star where most of the synthesized radioactive
elements fall back to a black hole.  With the discovery of Si~II and
S~II in the early-time spectra of 08ha \citep{foley10}, that
proposition becomes less likely.

Here, we report photometry and spectroscopy of SN~2007qd (hereafter
07qd). Its observed properties place it in the 02cx subclass of
supernovae, specifically as a member intermediate to 02cx and 08ha,
linking these objects.  We present the photometric and spectroscopic
observations of 07qd and compare its unique
properties with a range of other SN.  We estimate the rise time,
luminosity distance, and peak luminosity of this peculiar object, and
we calculate its kinetic energy.  Furthermore, we model the spectra to
determine the velocity and composition of the ejecta and compare these
with the findings of other similar SN to identify common traits.  With
this information we predict the synthesized $^{56}$Ni mass and
determine whether 07qd shares more in common with a thermonuclear
deflagration or a core-collapse scenario.

\section{Observations}
\subsection{Photometry}

\begin{figure}[h]
\epsscale{0.7}
\plotone{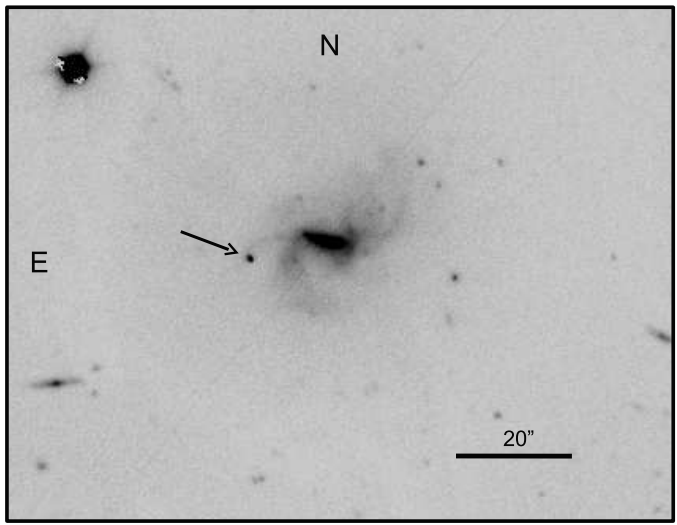}
\caption{\label{f:finderchart} Image of 07qd (denoted by arrow)
  relative to its host galaxy in a $105'' \times 80''$ window.  The
  90-s unfiltered exposure was taken on MJD 54409.08 with the TNG
  telescope.}
\end{figure}

07qd \citep{bassett07} was discovered during the SDSS-II Supernova
Survey using the SDSS Camera on the 2.5~m telescope
\citep{gunn98,gunn06} at Apache Point Observatory.  The SN was located
amidst a spiral arm at $\alpha = 02^h 09^m 33.56^s$, $\delta =
-01^{\circ} 00' 02.2''$ (J2000.0), a projected distance of 10.6 kpc
from the nucleus of the SBb/SBc host galaxy SDSS J020932.73-005959.8
centered at $\alpha = 02^h 09^m 32.73^s$, $\delta = -00^{\circ} 59'
59.80''$ \citep{bassett07}.  The redshift of the host is $z =0.043147
\pm 0.00004$, as measured from the SDSS galaxy redshift survey
\citep{york00,adelman08}.  Figure~\ref{f:finderchart} shows an image
of 07qd and its location in its host galaxy. The foreground Milky Way
extinction in the direction of 07qd is $E(B-V) = 0.035$ mag as
calculated from the dust maps of \citet{schlegel98}.

\begin{figure}
\epsscale{1}
\plotone{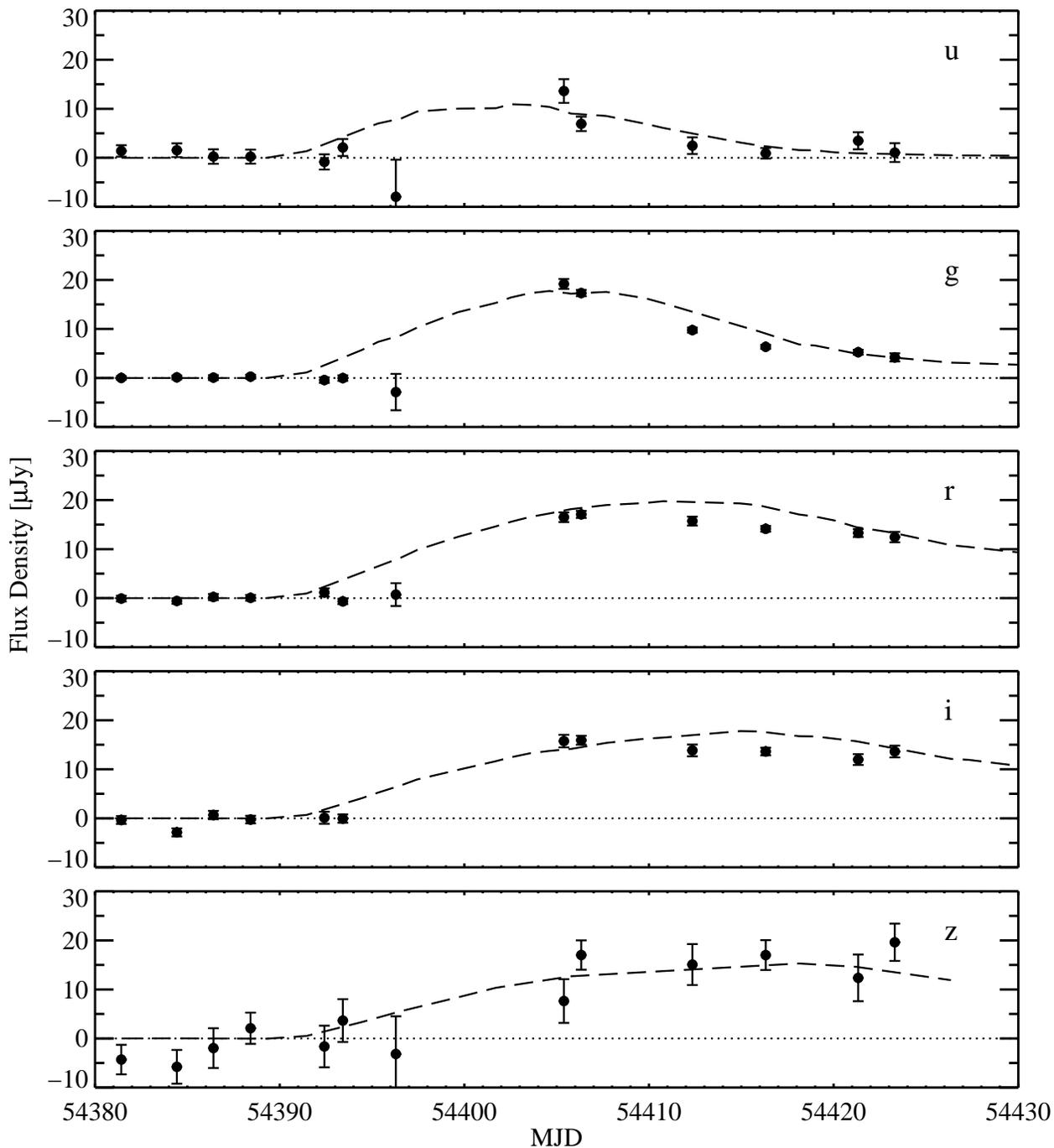}
\caption{\label{f:lightcurves2} SDSS apparent light curves of 07qd
  given in flux-density units.  The dashed line represents the
  combined SDSS and CSP light-curve data of 05hk in the observed frame
  of 07qd, positioned over the maximum brightness of 07qd, and
  artificially shown 2.4 mag fainter for better comparison of their
  shapes.  With respect to the early nondetections, 07qd clearly rose
  faster than 05hk in the $u$, $g$, $r$, and $i$ bands.  Neither 
  light curve has been corrected for host-galaxy extinction.}
\end{figure}

The photometry was calibrated in the standard SDSS $ugriz$ photometric
system \citep{fukugita96,smith02}.  The flux from the supernova was
estimated by using the scene modeling technique \citep{holtzman08}
from individual calibrated images and without spatial resampling.
Figure~\ref{f:lightcurves2} shows the SDSS-II light curves in the
$ugriz$ bands; the data are listed in Table~\ref{t:phot_data}.  The
time of peak bolometric flux for 07qd is not well defined, but probably occurred
within a span of 2 days around maximum apparent $g$-band magnitude
$20.69\pm0.06$ mag on MJD 54405.39.  We base this on both the available $u$, $g$, 
and $r$ data detected on that date, and the fact that the $u$ 
and $g$ fluxes were falling while the $r$, $i$, and $z$ fluxes were 
still rising.

\citet{bassett07} noted that the spectrum of 07qd was similar to that
of 05hk, so we compared their light curves.  While the maximum
$g$-band apparent magnitude of 05hk was $g = 16.32 \pm 0.02$, it
achieved a peak $B$-band absolute magnitude of $M_B=-18.0\pm0.3$.
This is rather dim for a typical SN~Ia ($M_B \approx -19$ mag), but
$\sim 2.6$ mag brighter than 07qd (see \S~3.1).  Featured in
Figure~\ref{f:lightcurves2} are the 05hk light curves corrected to the
observed frame of 07qd.  In addition, the brightness of 05hk has been
reduced by 2.4 mag to better match the peak flux of 07qd and enable
their shapes to be compared.  Ignoring possible differences in dust
extinction, 05hk is more luminous than 07qd in all of the SDSS bands,
and 05hk both rose and declined more slowly in $u$, $g$, and $r$ than
07qd. The light-curve widths of the two supernovae are similar at
near-infrared wavelengths (SDSS $i$, $z$), though 07qd shows a very
slow decline rate at red wavelengths. 05hk also appears to peak later than 07qd as the bandpass becomes progressively redder.

\begin{figure}[h]
\epsscale{0.7}
\plotone{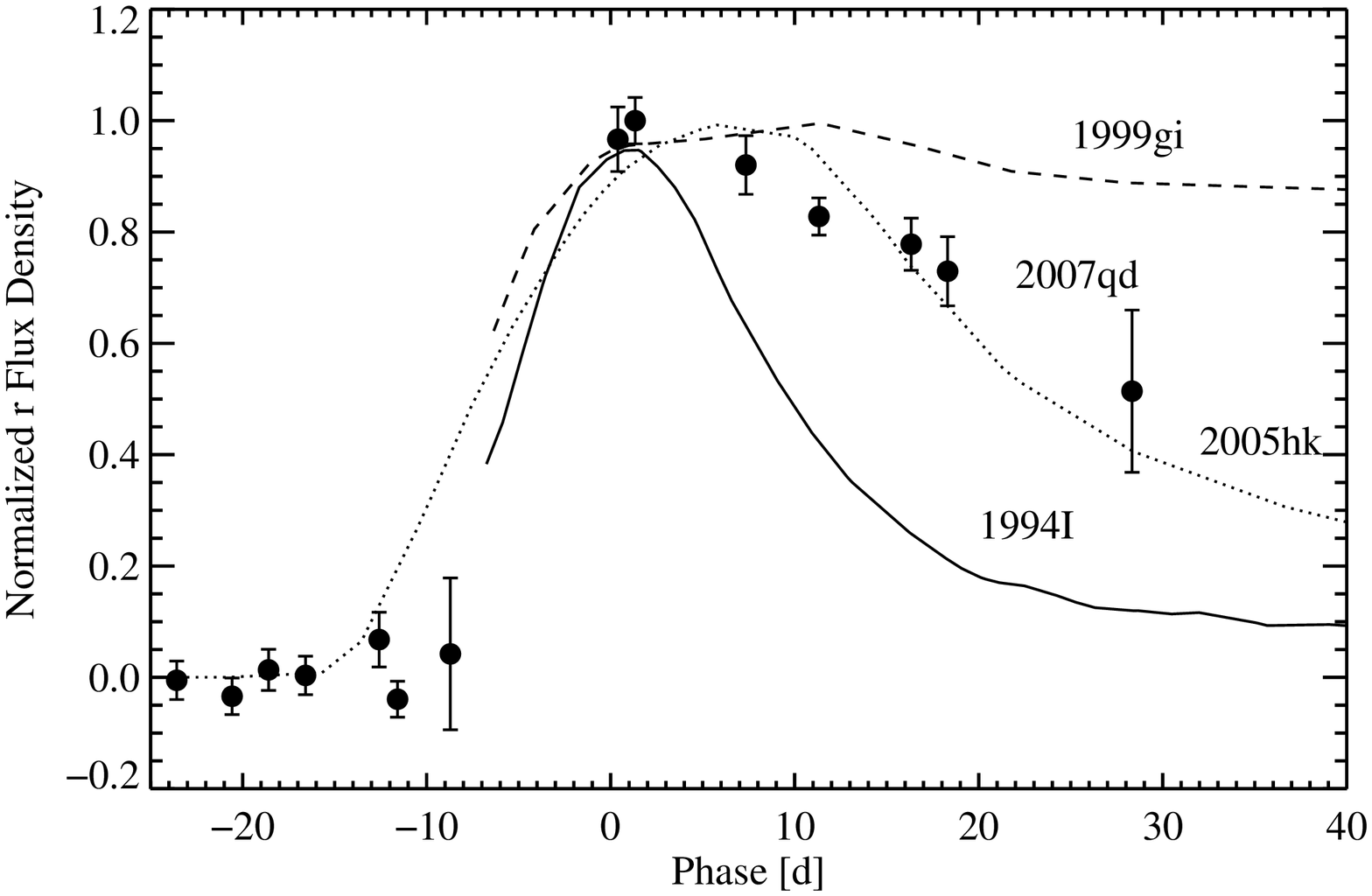}
\caption{\label{f:sn_types} $r$-band light curves of SN~1994I (Type
  Ic, solid line), SN~1999gi (Type IIP, dashed line), 05hk (dotted
  line), and 07qd (data points) given in flux units. SN~1994I and
  SN~1999gi have been converted to the SDSS $r$ bandpass via
  conversions utilizing $V$ and $R$ by \citet{fukugita96}, and all
  fluxes have been normalized to each maximum.  The time axis has also
  been corrected to the rest frame, and each SN arbitrarily
  shifted in phase to best match the rise portion.  SN~1994I,
  SN~1999gi, and 05hk are shown without error bars for clarity, and
  the interpolated segments were convolved with a Gaussian full width
  at half-maximum intensity (FWHM) of 2 days.}
\end{figure}

We compared the light curve of 07qd with those of additional
supernovae to identify any possibly similarity to the Type Ic or II
classes.  As 07qd rose rather quickly, we compared it to other events
which shared similar rise times.  We chose the fast-evolving SN~1994I
\citep{yokoo94,richmond96,sauer06}, a well-observed prototype for
SN~Ic not associated with a gamma-ray burst, and SN~1999gi
\citep{leonard02}, a Type IIP (plateau) SN whose absolute $B$
magnitude at peak was similar to that of 07qd (described in detail in
\S~3.1).  Figure~\ref{f:sn_types} exhibits the shapes of these light
curves in the $r$ band.  Though the steep rise implied by the 07qd
data may be similar to those of a SN~IIP or SN~Ic, the decline of 07qd
and 05hk fails to represent the steep drop of SN~1994I or the extended
plateau of SN~1999gi.  In addition, the rise of 07qd is far faster
than that of 05hk and is more similar to that of 08ha
\citep{foley09,foley10}.

\begin{figure}[h]
\epsscale{0.7}
\plotone{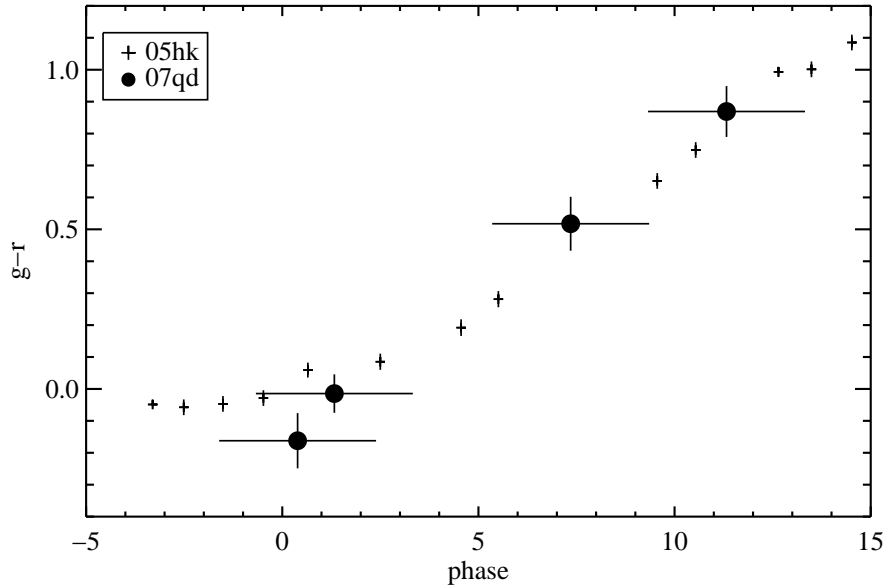}
\caption{\label{f:colorcomp_gr} The $g-r$ color of 05hk and 07qd for
  the first two weeks past maximum brightness.  The large uncertainty in
  the phase of 07qd stems from the 2-day uncertainty in the time of maximum.
  Neither SN has been corrected for reddening.}
\end{figure}

Figure \ref{f:colorcomp_gr} compares the color evolution of 07qd to
that of 05hk, and reveals the former to be more blue near maximum
brightness.  The colors of normal SN~Ia are fairly well established
\citep{phillips99, riess96} and are used to estimate the reddening
caused by dust in the host galaxy. However, the intrinsic colors of
peculiar 02cx-like events are uncertain, making the estimation of
host-galaxy extinction problematic. 07qd is bluer than 05hk, having
lower $g-r$ colors of $0.21\pm 0.09$ mag and $0.13\pm 0.06$ mag at 0.4
and 1.3 days past $B$ maximum, respectively.  This suggests that dust extinction 
is not the major cause of the low luminosity of 07qd compared with 05hk.
Additionally, 07qd appears to decline slower in the $i$ and $z$ light
curves.

\subsection{Spectroscopy}

The 3.58 m Telescopio Nazionale Galileo (TNG) in the Canary Islands
observed 07qd at 3 days after $B$ maximum (see \S 3.1 for the
determination of $B_{\mathrm{max}}$).  The spectrum is a composite of
three 30-min exposures obtained with DOLORES (Device Optimized for LOw
RESolution), a low-resolution spectrograph and camera permanently
installed at the Nasmyth B focus of the TNG.  It is equipped with a
$2048 \times 2048$ pixel E2V 4240 thinned, back-illuminated,
deep-depleted, Astro-BB coated CCD with a pixel size of 13.5
$\mathrm{\mu}$m and a field of view of $8.6' \times 8.6'$ with a
$0.252''\, \mathrm{pixel}^{-1}$ scale.  The spectra were observed with
the low-resolution blue grism (LR-B, dispersion
2.52~\AA\ $\mathrm{pixel}^{-1}$), covering the 3673--7401~\AA\ range
at an airmass of 1.27. A slit of $1.0''$ width, equal to the average
seeing, was used for the observations and was aligned with the
supernova and galaxy core at a position angle of $-39.3^{\circ}$.

The Hobby-Eberly Telescope (HET) at the McDonald Observatory in Texas
collected spectra of 20-min exposure times utilizing the Marcario Low
Resolution Spectrograph \citep[LRS;][]{hill98} at 8 and 15 days past
maximum.  At the prime focus, the LRS employed a $0.235''
\mathrm{pixel}^{-1}$ plate scale with a $1''$ wide by $4'$ long slit
and covered the 4075--9586~\AA\ range, though low signal-to-noise
ratios severely limit visibility past 8000~\AA.

We obtained low-resolution spectra at 10 days past maximum brightness
with the Low Resolution Imaging Spectrometer \citep[LRIS;][]{oke95} on
the 10~m Keck~I telescope on Mauna Kea, Hawaii.  The Keck measurement
was able to cover bluer wavelengths than the other observations,
spanning the 3073--8800~\AA\ range.  For all spectra, standard CCD
processing and spectrum extraction were performed with
IRAF\footnote{IRAF: the Image Reduction and Analysis Facility is
  distributed by the National Optical Astronomy Observatory, which is
  operated by the Association of Universities for Research in
  Astronomy, Inc. (AURA) under cooperative agreement with the National
  Science Foundation (NSF).}.  The data were extracted using the
optimal algorithm of \citet{horne86}.  Low-order polynomial fits to
calibration-lamp spectra were used to establish the wavelength scale,
and small adjustments derived from night-sky lines in the object
frames were applied.  We employed our own IDL routines to flux
calibrate the data and remove telluric lines using the well-exposed
continua of spectrophotometric standard stars \citep{wade88, foley03}.

\begin{figure}[h]
\epsscale{0.8}
\plotone{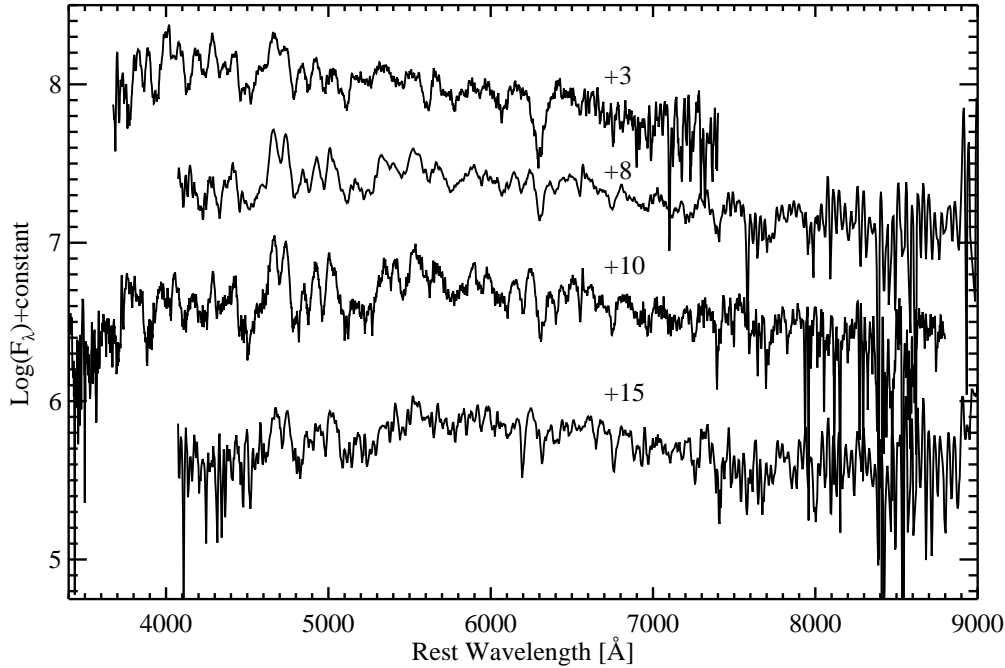}
\caption{\label{f:qd_evolve} Temporal evolution of the spectrum of
  07qd with a boxcar smoothing of 3 pixels in order to clearly show
  major features.  Times in this figure are given in days past
  $B$-band maximum, telluric absorption has been removed, and
  wavelengths have been corrected to the rest frame. No extinction
  corrections have been applied.}
\end{figure}

   A journal of the spectroscopic observations is given in
Table~\ref{t:spec_sched}, and the resulting extracted and reduced
spectra are detailed in Figure \ref{f:qd_evolve}.

\section{Analysis}
\subsection{Energetics}

We infer a rise time of $10 \pm 2$ days based on the nondetections in
the SDSS-II data, much shorter than that of typical SN~Ia
\citep{hayden10}.  We began our analysis by applying the light-curve
fitters MLCS2k2 \citep{jha07} and SALT2 \citep{guy07} to the 07qd,
05hk, 02cx, and 08ha data using the SNANA platform \citep{kessler09b}.
The resulting fits were of very poor quality, with $\chi^2 > 60$ per
degree of freedom, suggesting the inability of these algorithms,
trained on normal events, to fit the colors and light-curve shapes of
02cx-like peculiar supernovae.

We decided to take a simple approach and estimate the $B$-band
light-curve stretch \citep{goldhaber01} versus peak brightness for the
peculiar events and some more normal SN~Ia. We used SNANA
\citep{kessler09b} to convert from $ugriz$ to the standard $UBVRI$ for
SDSS-II SN~Ia with $z<0.12$, verifying the uncertainty with the
conversions of \citet{fukugita96}.  SNANA applies the MLCS2k2
K-correction algorithm, which is based on a color-matched normal SN~Ia
spectral energy distribution.  This may introduce small errors due to
differences in the spectral features between 07qd and typical SN~Ia.

For 07qd we find a maximum absolute magnitude (after correcting only
for Milky Way Galaxy extinction) of $M_B = -15.4 \pm 0.2$ mag.  We
estimate that the time of peak $B$ brightness occurred on MJD $54405
\pm 2$, though the true value could be smaller due to the lack of data
on the rise.  The slopes of the blue light curves of 07qd flatten out
rapidly after 10 days.  Consequently, $\Delta m_{15}(B) \approx 1.5$
mag for 07qd, or roughly the same as that of 05hk \citep{phillips07}.
As seen in Figure~\ref{f:lightcurves2}, 07qd clearly fades more
quickly than does 05hk, suggesting that in this case, the measurement
of $\Delta m_{15}(B)$ does not compare well with other SN~Ia.

\begin{figure}[h]
\epsscale{.7}
\plotone{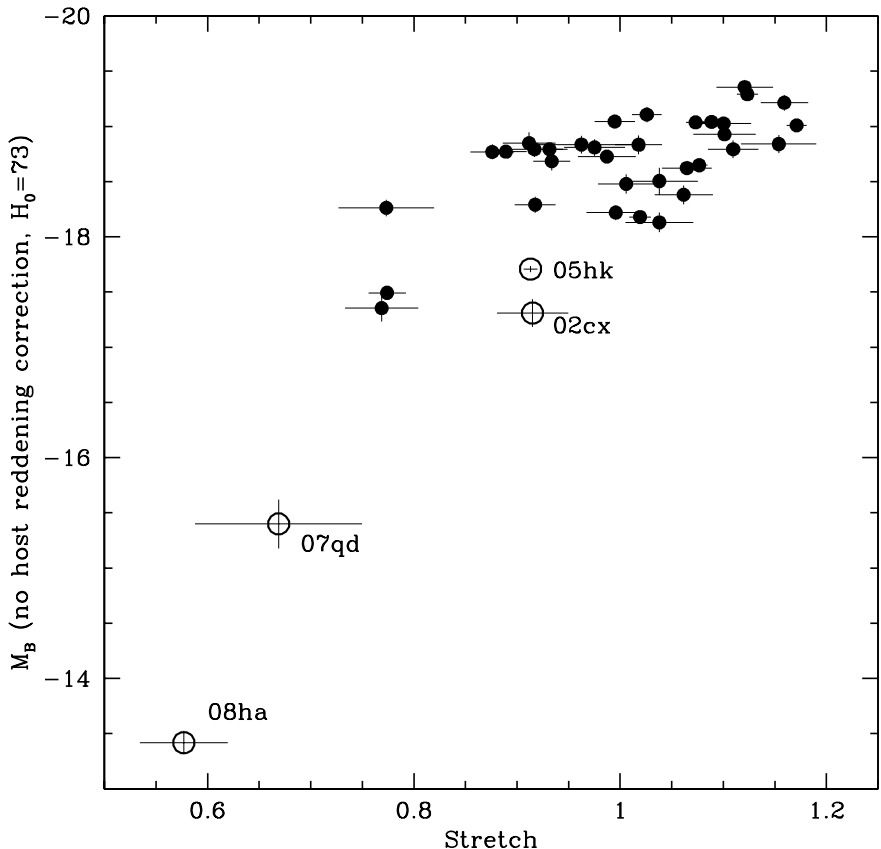}
\caption{\label{f:stretch73} Light-curve stretch factors are compared
  to the absolute magnitudes of peculiar SN~Ia (open circles) along
  with SDSS-II SN~Ia at $z < 0.12$ \citep{frieman08,sako08}.  Maximum
  $B$ for 07qd is based on an estimated 10-day rise time.  No
  correction for host-galaxy extinction has been made.}
\end{figure}

Based on our estimation of the rise time, Figure~\ref{f:stretch73}
compares the absolute magnitude of 07qd and its $B$-band stretch with
those of 02cx, 05hk, 08ha, and the normal SDSS-II SN~Ia (see
\citealt{sako08} for a list of these SN). The low-redshift set of
SDSS-II contains a handful of 91bg-like events with stretch parameters
$\sim 0.8$, but 07qd and 08ha have narrower light curves and are much
fainter.  There appears to be a sequence connecting the bright,
peculiar events 05hk and 02cx to the extreme 08ha with the
intermediate 07qd in Figure~\ref{f:stretch73}.

To examine the bolometric luminosities, a blackbody was fit to the
$ugriz$ spectral energy distribution (SED) as measured at $B$ maximum
brightness.  Ultraviolet measurements shortward of $u$ were not
available for 07qd. 05hk and 08ha experienced severe photospheric line
blanketing at $\lambda < 3500$~\AA\ \citep{phillips07, foley09},
which, along with our Keck spectrum (10 days past maximum), suggests
that the luminosity of 07qd drops off sharply at that point.  Using a
Hubble constant of $H_0 = 73$ km $\mathrm{s}^{-1}$ $\mathrm{Mpc}^{-1}$
\citep{freedman01,riess09} and our measured redshift, a luminosity distance of 177~Mpc ($m-M = 36.2$
mag) was computed.  Integrating our SED (corrected for
Milky Way extinction) and applying our luminosity distance, we 
calculated the peak quasi-bolometric luminosity to be $(4.4 \pm 0.5) \times 10^{41}$ ergs $\mathrm{s}^{-1}$.
Assuming ``Arnett's rule'' \citep{arnett82,arnett85} as applied by
\citet{stritzinger06}, we estimate the synthesized $^{56}$Ni mass
using the ratio of maximum luminosity to the luminosity produced by
one solar mass of $^{56}$Ni,

$$ M_\mathrm{Ni} = L_{\rm max} / E_{\rm Ni}(1{\rm M}_\sun) = \frac{ (4.4 \pm
  0.5)\times 10^{41}\, \mathrm{ergs\,\,s^{-1}} }{ \alpha(6.45 \, \mathrm{e}^{-t_{\rm R} \lambda_{\mathrm{\rm Ni}}} + 1.45 \, \mathrm{e}^{-t_{\rm R} \lambda_{\mathrm{\rm Co}}} ) \times 10^{43}\, \mathrm{ergs\,\, s^{-1}\, \,M_{\sun}^{-1}}}, $$

\noindent
where $\lambda_{\mathrm{Ni}}=1/8.8$ days and
$\lambda_{\mathrm{Co}}=1/111.3$ days are the e-folding times for
$^{56}\mathrm{Ni}$ and $^{56}\mathrm{Co}$, respectively.  Thus, for
our maximum possible
$t_{\rm R} = 10 \pm 2$ day rise time, we estimate an upper limit of $M_\mathrm{Ni} = 0.013 \pm
0.002$ M$_{\sun}$, which is nearly a factor of 5 larger than that of
08ha [$(3.0 \pm 0.9 )\times 10^{-3}$ M$_{\sun}$; \citealt{foley09}],
but still much smaller than that of 05hk [$\sim 0.22\,\, {\rm M}_{\sun}$;
  \citealt{phillips07}].  Here, we assume that the efficiency of
radioactive energy deposition $\alpha$ is unity (see discussion in 
\citealt{howell09}).

The ejecta mass may be estimated by utilizing the treatments of
\citet{arnett82} and \citet{pinto00a,pinto00b}, assuming an opacity of
0.1 $\mathrm{cm^{2}}$ $\mathrm{g^{-1}}$, as per an Fe~II spectrum in
the same vein as \citet{foley09}.  Comparing 07qd to a typical SN~Ia
of kinetic energy $10^{51}$ ergs, rise time 17 days \citep{hayden10}, 
and photospheric velocity 10,000 km s$^{-1}$, we derived an ejecta mass of
0.15~M$_\odot$.  This could be even lower if the rise time of 07qd 
is under 10 days or using higher estimates of standard SN~Ia rise times.
Based on a photospheric velocity 3 days past maximum brightness
of 2800 km s$^{-1}$ (see \S 3.2.1), this translates to a kinetic
energy of $\sim 2.0 \times 10^{49}$ ergs.  This is similar to the
kinetic-energy estimate of 08ha by \citet{foley10}, but roughly an
order of magnitude lower than the $\sim 8 \times 10^{50}$ ergs of 05hk
based on its photospheric velocity \citep{phillips07}.  Typical SN~Ia
contain $^{56}$Ni masses between 0.1 and 1.0 M$_\odot$ and ejecta
masses ranging from 0.5 M$_\odot$ for SN~1991bg to the
canonical 1.4~M$_\odot$ \citep{stritzinger06}.  Both of these values
are significantly larger than our findings for 07qd.

\subsection{Spectroscopy}

We used SYNOW \citep{fisher97, fisher00}, a parameterized supernova
spectrum synthesis code, to fit our spectra to profiles of various
ions at specific velocities, excitation temperatures, and opacities.
These ions are simulated in an expanding photosphere of a chosen
blackbody temperature, and the resulting spectra can be compared with
data.  We systematically fit our spectra with ions commonly seen in
SN~Ia at similar epochs (Fe~II, Co~II, Si~II; see \citealt{hatano99}
and \citealt{maeda10} for lists of expected ions and isotopic yields),
and attempted to simulate hydrogen and helium to rule out the
possibility of a Type II or Ib SN.  The photospheric velocity was
fixed at each epoch by the best fit for the Fe~II lines, and
subsequent elements were fit either residing in that photosphere or at
detached velocities.  After confirming the presence or absence of
these species, we tried to fit elements uncommon to normal SN~Ia in
order to fit any remaining line profiles.

We also attempted to use Na~I~D host and interstellar absorption to
measure dust extinction though the procedures of \citet{munari97}.
The low signal-to-noise ratio of our spectra, combined with features
at various velocities scattered throughout the continuum, created too
much uncertainty in these equivalent widths to fix an extinction
value.

\subsubsection{+3 Days Spectrum}

\begin{figure}
\epsscale{1}
\plotone{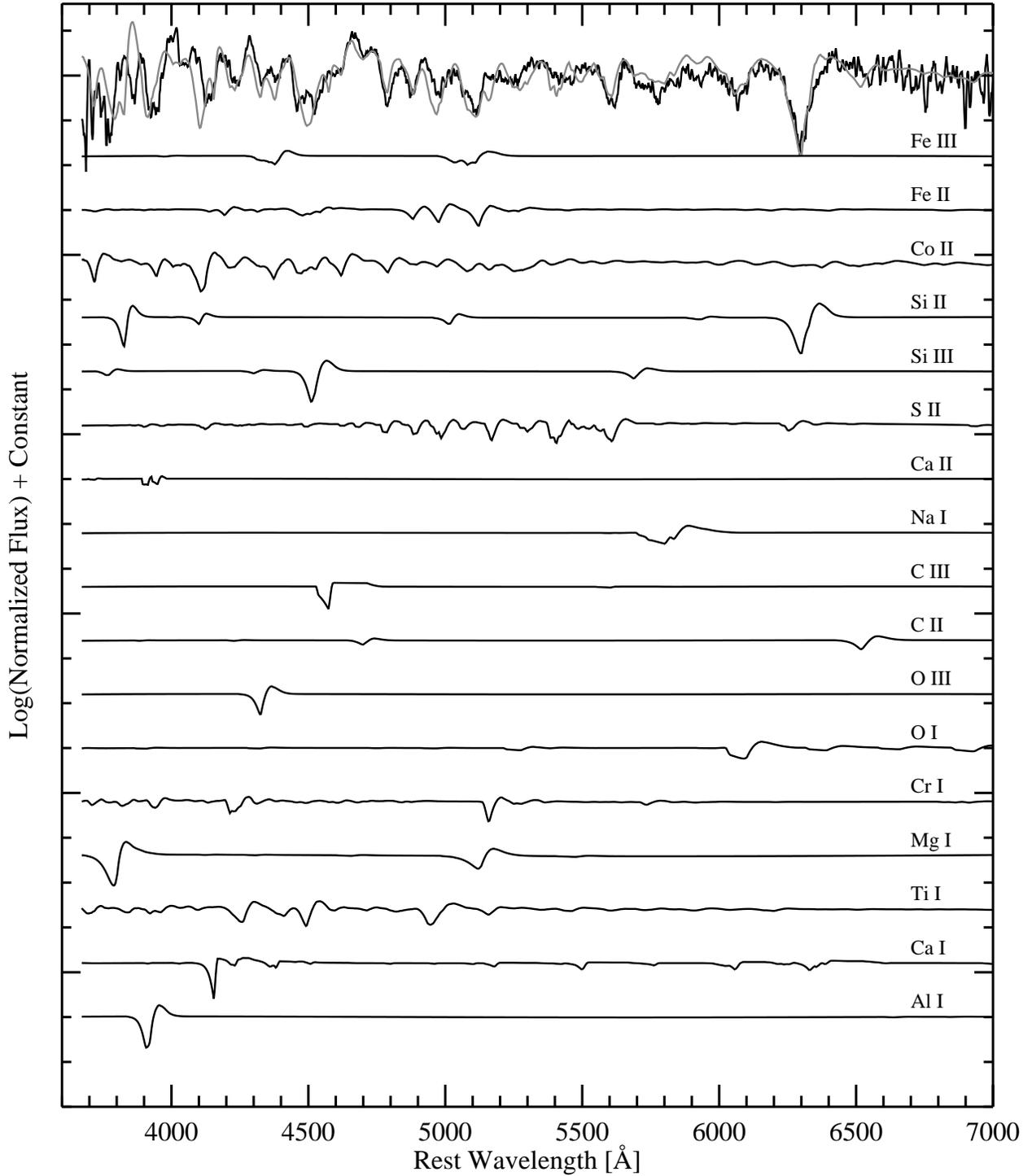}
\caption{\label{f:3d_decomposed} The spectrum of 07qd at 3 days past
  maximum (top) and the best SYNOW fit (gray line). The contribution
  of each individual species is also shown.  All spectra have been
  normalized, while the data and combined fit have been scaled up by a
  factor of 2 for clarity.  Our more confident identifications are presented
  near the top while lines at the bottom are less likely.}
\end{figure}

Figure \ref{f:3d_decomposed} and Table~\ref{t:qd_3d} present the
results of our SYNOW fit to the spectrum 3 days after maximum
brightness, as well as a decomposition of the fit to show the
influence of each ion.  We found from the Fe~II lines a photospheric
velocity of 2800~km $\mathrm{s^{-1}}$, which is extraordinarily
low. Typical SN~Ia photospheric velocities at this epoch are often in
excess of 10,000 km $\mathrm{s}^{-1}$ \citep{pskovskii77, branch81}.
The low velocities found for most of the regions makes it highly
unlikely that the strong feature at 6300~\AA\ is from hydrogen or
helium either in the photosphere or detached.  The best fit to the
6300~\AA\ absorption was Si~II, since the velocity of Fe~II was
constrained when fitting the $\sim 6100$~\AA, $\sim 6200$~\AA, and
$\sim 6400$~\AA\ features as well as others in the bluer spectral
regions.  Si~II fit the absorption best when at a velocity of 800 km
s$^{-1}$ lower than that of the iron-group ions.  Also prominent in
the $\sim6100$~\AA\ region is a broad primary O~I line.  The secondary
signatures of O~I are weakly detected at $\sim 5300$~\AA, but masked
at $\sim 6400$~\AA\ due to the strong Si~II feature nearby.

Si~III and C~III were implemented in a manner similar to that used by
\citet{chornock06} to help Co~II shape the area around $\sim
4600$~\AA.  C~II, also present in the maximum-light spectrum of 08ha
\citep{foley10}, was necessary for an absorption feature at
$\sim6550$~\AA.  Our SYNOW fits suggest the presence of Mg~I, Ti~I,
Cr~I, Ca~II, and Al~I.  \citet{hatano99} do not predict these species
in a thermonuclear explosion when Si~III is strong.  It is likely that
the features we attribute to these ions are due to other unidentified
species or different velocities for existing ones, though we cannot
definitively rule out these unusual identifications.  It should also
be noted that C~I, O~II, Ni~II, and Co~III, predicted by
\citet{hatano99}, may be added with little effect on the overall fit.
We were unable to fit the absorption between 3800 and 3900 \AA\, at
this photospheric velocity with anything other than K~II, though its
inclusion would introduce other discrepancies to the fit.

\subsubsection{+10 Days Spectrum}

\begin{figure}
\epsscale{1}
\plotone{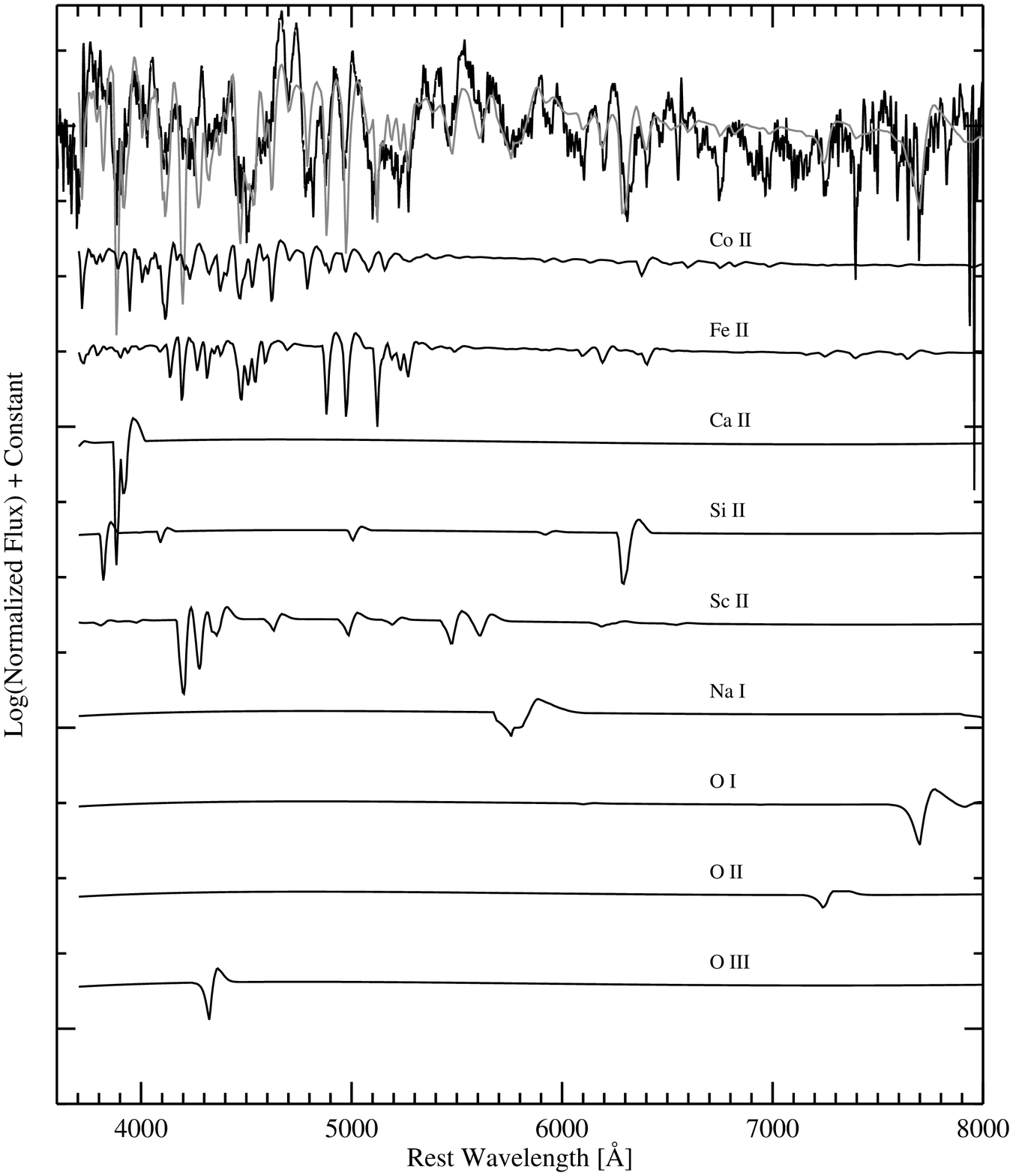}
\caption{\label{f:qd_synow_10d} Normalized 07qd spectrum 10 days after
  maximum, with SYNOW fit and its constituents.  Again, the data and
  combined fit have been scaled up by a factor of 2 for clarity, and
  species are listed roughly in order of decreasing confidence.}
\end{figure}

\begin{figure}[h]
\epsscale{.7}
\plotone{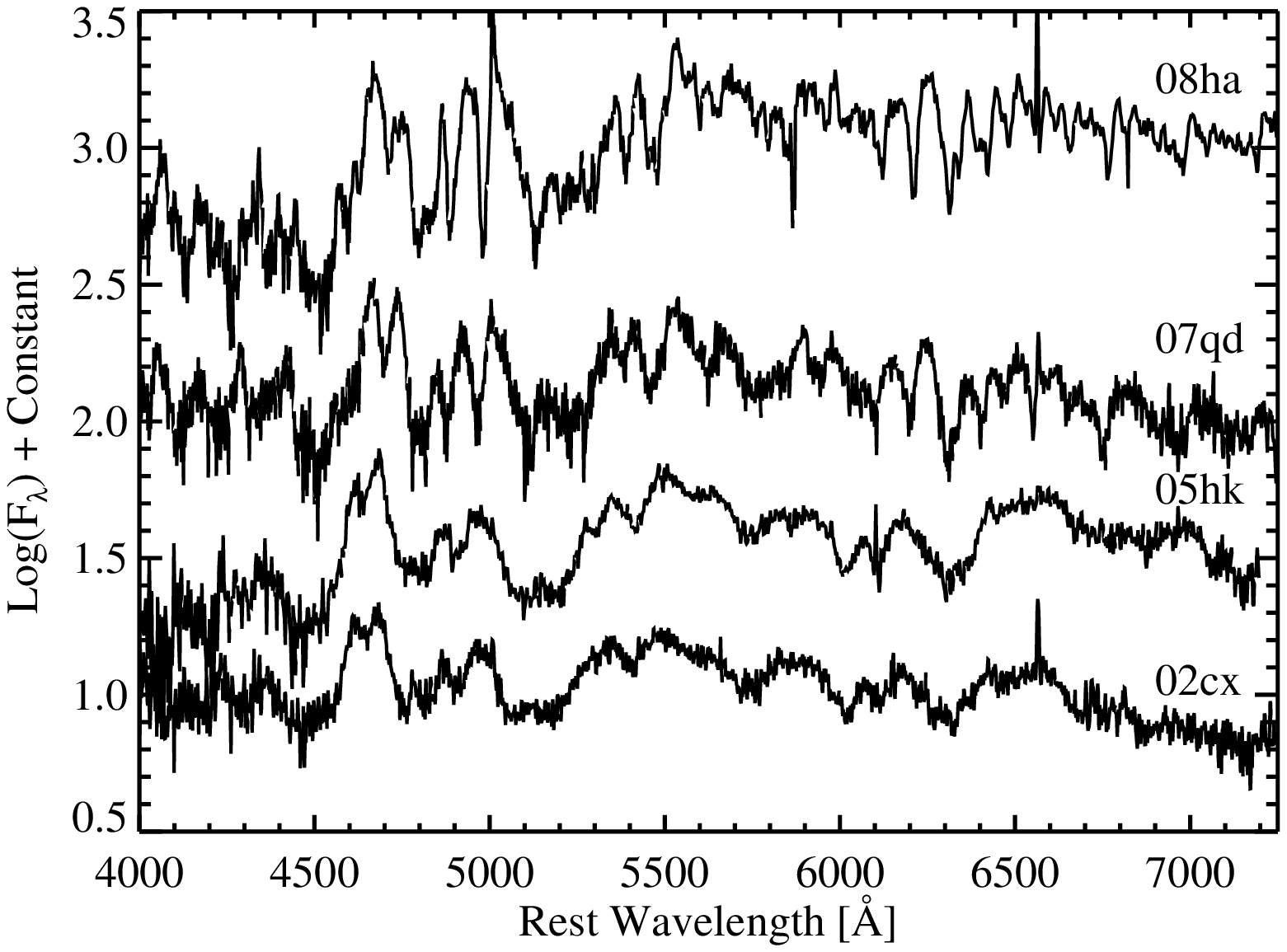}
\caption{\label{f:qd_hk_cx} Comparison of the spectrum of 07qd at 10
  days past maximum brightness with that of 08ha \citep{foley09}, 05hk
  \citep{phillips07}, and 02cx \citep{branch04} at similar epochs (10
  days for 08ha, and 12 days for both 05hk and 02cx).  All spectra
  have been appropriately shifted to their rest frames.  Each feature
  of 07qd is redward of those of 05hk and 02cx, indicating
  significantly slower photospheric velocities, while only slightly
  blueward of those of 08ha.}
\end{figure}

The SYNOW fit in Figure \ref{f:qd_synow_10d} is derived from
parameters used to fit 02cx at 12 days past maximum \citep{branch04},
except the velocities and blackbody temperature have been reduced to
accommodate the unique nature of 07qd.  This fit was chosen due to its
consistency in representing 08ha, 02cx, and 05hk at similar epochs, as
is seen in Figure \ref{f:qd_hk_cx}.  The similarity is especially
apparent with 08ha, which shares most absorption features with 07qd.
The relative shift between these two spectra was measured to be $\sim
800$ km $\mathrm{s^{-1}}$; the photospheric velocity of 08ha at this
epoch is 2000 km s$^{-1}$, while the best-fit SYNOW photospheric
velocity of 07qd is $\sim 2800$ km s$^{-1}$ with a blackbody
temperature of 8000~K.  Though \citet{foley09} found an excellent
match to the 14-day spectrum of 08ha with a low photospheric velocity
of only 600 km s$^{-1}$, we attempted our fits with the Fe~II
estimated velocity of 2000 km s$^{-1}$.

Much of the continuum is dominated by narrow Fe~II lines, with Co~II
playing a large role below 5000 \AA\ as well, consistent with most
iron-core SN \citep{harkness91}.  The modified SYNOW fit also
continues to include a prominent Si~II feature at
6300~\AA\ necessitated by the inability of Fe~II to fit it without
losing other features.  This Si~II resides within the photosphere and
enhances the spectrum at $\sim 4100$~\AA, but less so at
$\sim3850$~\AA\ and $\sim5025$~\AA.  The carbon and vanadium measured
at 3 days past maximum are not prominent at any ionization stage.
Their existence cannot be ruled out, in part due to the influence of
the dominant Fe~II lines \citep{baron03}.  The Cr~II in the 02cx fit
by \cite{branch04} is also not necessary, as there is insufficient
data to examine regions below 3750~\AA\ at this and later epochs.  We
were unable to identify any ionization stage of Ti at this epoch,
despite the possibility of observing Ti~I at 3 days past maximum.
\citet{foley09} identified Ti~II at the same epoch in 08ha, but its
addition here does little to the overall fit.  Other ions include
Sc~II (for the absorption at 5450~\AA) and Ca~II (for the H\&K
lines). The secondary absorption lines S~II are no longer apparent at this epoch, as Sc~II is able to
effectively represent the 5400--5700~\AA\ region without aid.

The signal-to-noise ratio of the data appears to be lower than it
actually is, due to the low-velocity nature of this particular
photosphere.  Much of the absorption blending in the regions blueward
of 4500~\AA\ proved difficult to fit with SYNOW models, especially
where P-Cygni profiles are no longer apparent; it is suspected to be a
consequence of the narrow absorption features of several iron-group
elements.

H$\alpha$ and helium ions were tested in SYNOW fits, though no match
could be found.  We expect and observe host-galaxy H$\alpha$ emission at
6563~\AA, though just blueward of this feature we see a small
absorption that we identify with C~II.  The nature of the trough at
$\sim6750$~\AA\ is unclear at this time; very few ions are capable of
modeling it without severely affecting the fit elsewhere.  He~I was of
particular interest in the SYNOW fit to the 08ha spectrum 13 days past
maximum conducted by \citet{valenti09} and to SN~2007J, another
02cx-like SN \citep{foley09}, but the Na~I~D line dominates over the
predicted He~I line.  At 10 days past maximum, all other He~I features
fail to match the spectrum.

\subsubsection{Spectral Evolution and Similarity to Other SN~Ia}

Using these SYNOW fits to the individual epochs, the spectral
evolution of 07qd given in Figure~\ref{f:qd_evolve} provides a
detailed picture of a developing photosphere.  Examining the
contributions near $\sim4550$ and $\sim5200$~\AA\ present at 3 days
but absent at 10 days, Fe~III has either greatly fallen in opacity
or recombined into Fe~II, which has increased in influence.  Na~I and
S~II have also decreased in intensity, but remain crucial to the
region between 5000 and 6000~\AA.

The Si~II feature endures through two weeks past maximum, but its
strength has lessened, becoming roughly equal to that of Fe~II by day
15.  Much of the spectrum in the blue region was not measured at
subsequent epochs, though extended red wavelengths are given,
revealing probable O~I and O~II signatures.  It is apparent, however,
that the early-time spectra beyond 8000~\AA\ demonstrate the presence
of the Ca~II near-IR feature, though the Ca~II H\&K near-UV lines
persist.  Blackbody temperatures have fallen to 8000~K at 8 days past
maximum and to 6000~K at 13 days, further intensifying the influence
of Fe~II profiles over the continuum.  It remains to be seen whether
the other unidentified line profiles can be remedied with more exotic
ions.

Figure \ref{f:qd_hk_cx} shows the spectra of four 02cx-like SN~Ia
compared at similar epochs.  07qd clearly bears the strongest
resemblance to 08ha; very few features differ.  02cx and 05hk exhibit
much faster photospheres than 07qd and 08ha, and are likewise shifted
toward the blue.  The faster photosphere and higher excitation
temperatures used in the SYNOW fit of 02cx \citep{branch04} broaden
many of the Fe~II lines, effectively masking the primary Si~II
absorption, but other major velocity features remain consistent.  It
should be noted, however, that SYNOW's highly parameterized fitting
routine permits many degrees of freedom.

\begin{figure}[h]
\epsscale{.7}
\plotone{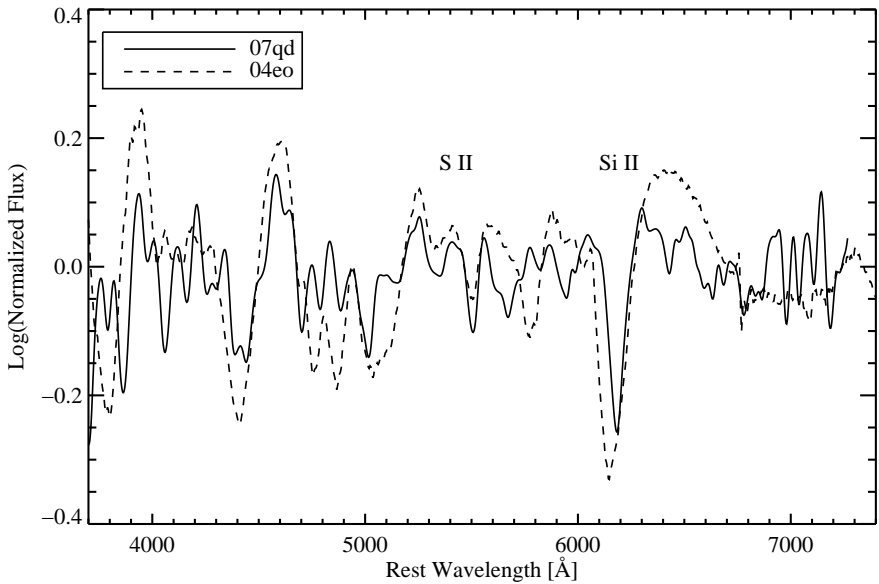}
\caption{\label{f:qdvseo} Normalized spectrum of 07qd compared with
  that of SN~Ia 2004eo \citep{pastorello07} and shown in the rest
  frame.  The day 3 spectrum of 07qd has been convolved with a
  30~\AA\ FWHM Gaussian and blueshifted by 5500 km $\mathrm{s}^{-1}$ 
  to emphasize the similar Si~II and S~II features.}
\end{figure}

In order to independently identify these IMEs outside of models, we
directly compared spectra of 07qd with those of other confirmed SN~Ia.
Of special interest is the spectral relation of 07qd to SN~2004eo
\citep[hereafter 04eo;][]{pastorello07}.  Similar to the prototypical
SN~1992A \citep{hamuy96c}, 04eo is characterized as a fast-declining
SN~Ia with slow photospheric velocities.  However, it retains an
absolute $B$ maximum of $-19.08$ mag (within the range of ``normal''
SN~Ia) and still fits the Phillips (1993) relation.
Spectroscopically, 04eo contains strong Si~II and S~II lines, typical
of SN~Ia.  When the spectra of 07qd are blueshifted to coincide with
the photospheric velocity of 04eo, many common features become
apparent, especially the Si~II and S~II lines.  Figure \ref{f:qdvseo}
shows a normalized comparison between early-time spectra of 07qd and
04eo that highlights both the presence and the relative strengths of
these lines.

\begin{figure}[h]
\epsscale{.7}
\plotone{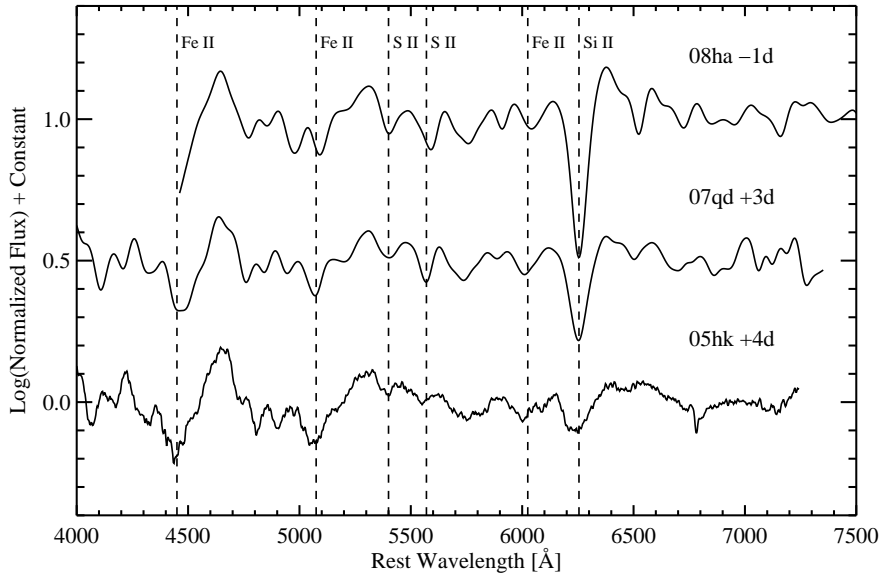}
\caption{\label{f:qdvshk_3d} Normalized spectra of peculiar SN~Ia,
  observed near maximum brightness.  The spectra of 07qd and 08ha have
  been blueshifted by 2000 and 1750 km s$^{-1}$, respectively, to
  roughly match the photospheric velocity of 05hk.  07qd and 08ha have
  also been convolved by 50~\AA\ FWHM Gaussians, while 05hk was smoothed
  with a 5~\AA\ FWHM Gaussian.}
\end{figure}

Similarly, we may compare the maximum-light spectra of 07qd, 08ha, and
05hk, as seen in Figure~\ref{f:qdvshk_3d}.  The spectra of both 07qd
and 08ha, with photospheres blueshifted by 2000 km~s$^{-1}$ and 1750
km~s$^{-1}$ (respectively), mirror many of the major features.
Although the Si~II and S~II profiles are noticeably stronger in 07qd
and 08ha, their presence in 05hk is clear.  The similarity between
07qd and 05hk below 5000~\AA\ is striking --- identical structures in
this region are indicative of similar iron-group contributions.

\begin{figure}[h!]
\epsscale{.7}
\plotone{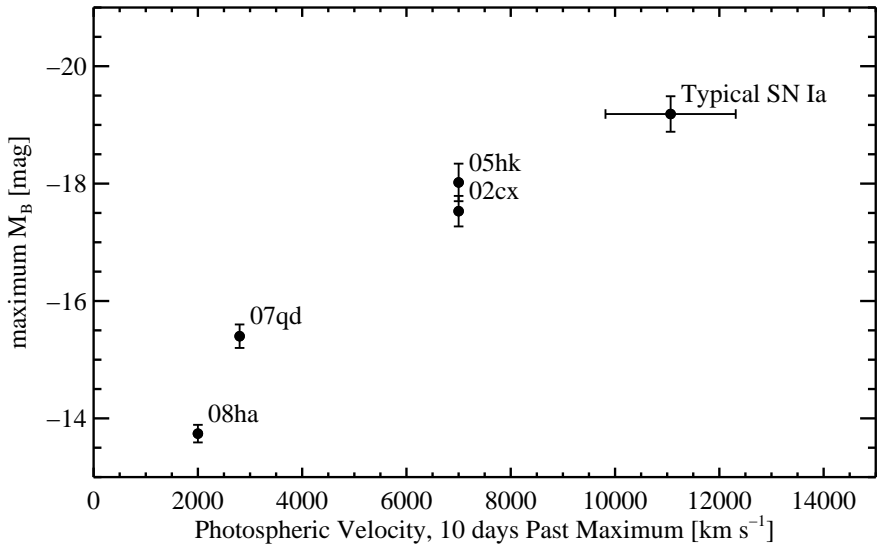}
\caption{\label{f:vphot_Bmax} Estimated photospheric velocities at
  $\sim 10$ days past maximum brightness are referenced by their
  maximum absolute magnitude (based on a $\sim10$ day rise time for
  07qd).  The ``typical'' point in the upper-right corner is the
  root-mean square maximum $B$ magnitude and velocity of the normal
  SN~Ia found by \citet{benetti05}.}
\end{figure}

The greatly different photospheric velocities apparent in 07qd and
05hk are examined along with 02cx and 08ha in Figure
\ref{f:vphot_Bmax}, suggesting a relation between the luminosities and
expansion velocity in these peculiar cases.  We used the Fourier
cross-correlation subroutine (FXCOR) in IRAF to determine the average
offset of the similar features in these spectra, namely the Fe~II and
Si~II lines. The average photospheric velocities of 07qd are faster
than those of 08ha at all epochs when both have measured spectra, but
lower than those of any other 02cx-like event.  We compare at 10--12
days past maximum because it is here that we have spectra of all four
of these 02cx-like SN. To first order, we extrapolated the
photospheric velocity of 07qd at maximum to be roughly 200 km s$^{-1}$
greater than that of 08ha, implying a larger kinetic energy by a
factor of 1.14 (assuming the same ejected masses for both SN).  Since
our calculation in \S~3.1 assumes nickel mass is directly related to
luminosity and varies with the rise time by a $\sim 7\%$ change in
$^{56}$Ni mass per day, the rise-time dependence is small.  As a
result, the plot of the photospheric velocity versus peak luminosity
suggests that the $^{56}$Ni yield (via maximum brightness and rise
time) and kinetic energy per unit mass (via photospheric velocity) are
correlated in 02cx-like events.

Through FXCOR, the average rate of decline in the photospheric
velocity is found to be $93 \pm 15$ km s$^{-1}$ day$^{-1}$ for the
6355~\AA\ Si~II line and $70\pm 21$ km s$^{-1}$ day$^{-1}$ for the
other absorption features (dominated by Fe~II).  Both of these values
group the velocity evolution of 07qd amongst the faint SN~Ia described
by \citet{benetti05}, and they are smaller in magnitude than those of
08ha ($142\pm26$ km s$^{-1}$ day$^{-1}$ for Si~II and $110\pm10$ km
s$^{-1}$ day$^{-1}$ for all other ions).  It is difficult to determine
the Si~II velocities after maximum brightness for 05hk and 02cx since
they are quickly blended with Fe~II lines.  \citet{phillips07} found
for both 05hk and 02cx that the velocities of the 4555~\AA\ Fe~II line
and the Ca~II H\&K lines remained constant at roughly 6000 km s$^{-1}$
for several weeks, implying that the overall expansions were quite
slow even for typical SN~Ia.  If we assume a constant deceleration
through late times, the 700 km s$^{-1}$ Fe~II lines found 227 days
past maximum in 02cx imply a 23 km s$^{-1}$ day$^{-1}$ decline
\citep{jha06}.

\subsection{Host Galaxy}

The host galaxy in Figure \ref{f:finderchart} shows a prominent bar
and two major spiral arms.  The low inclination allows minor spiral
arms to be distinguishable, one of which contains the location of
07qd.  The classifications of SBb or SBc suggest ongoing star
formation, but 02cx-like SN~Ia appear to span a wide variety of
galactic morphologies \citep{foley09,valenti09}.  The Keck-I spectrum
also included the host galaxy's core in the long slit, allowing a
measurement of the oxygen abundance.  We used both the optimized
method recommended in \cite{kewley02} and the
$[\mathrm{N~II}]/[\mathrm{O~II}]$ diagnostic to arrive at an oxygen
abundance in the host galaxy of
$12+\log{(\mathrm{O}/\mathrm{H})}=8.79\pm0.03$.  This is equivalent to
the Sun's oxygen abundance of $8.76 \pm 0.07$ \citep{caffau08}, but it
exceeds that of the host galaxy of 08ha, measured to be $8.16 \pm
0.15$ by \cite{foley09} using N2 and O3N2 diagnostics (see
\citealt{pettini04}).

\section{Discussion}

The striking similarity of the +10 day (relative to $B$ maximum)
spectrum of 07qd to that of 08ha suggest that they are indeed similar
explosions.  Since 07qd is also spectroscopically linked to 05hk and
thus to 02cx as well (see Figure~\ref{f:qdvshk_3d}), these four
peculiar events range in peak luminosity by 4 mag, but constitute a
single spectral class.

\cite{valenti09} argued that 08ha and possibly other 02cx-like objects
are actually core-collapse SN; their low luminosity is the result of
either the collapse of a $\gtrsim 30$ M$_\sun$ star directly to a
black hole, or electron capture in the core of a 7--8 M$_\sun$ star.
However, the clear presence of IMEs (notably Si~II $\lambda$6355 and
S~II $\lambda\lambda$5968, 6359) in 07qd suggests low-density
thermonuclear burning and not a core-collapse SN~Ib or SN~Ic explosion
(see \citealt{filippenko97} for a review of SN spectroscopic
classification).  Except for the photospheric velocities, the spectra
of 07qd are similar to those of SN 1992A-like SN~Ia, further
strengthening the arguments that it is a thermonuclear runaway.

The presence of these IMEs mirrors the findings of \citet{foley10}.
SN~Ibc have been observed with traces of Si~II \citep{valenti08} or
S~II \citep{nomoto00,brown07}.  However, these lines are quite strong
in 07qd, suggesting that it is a SN~Ia rather than a core-collapse
event.  Additionally, the light curve of 07qd, though displaying a
sharp rise that is unusual for a typical SN~Ia, does not show the fast
decline typical of a low-luminosity SN~Ic or the ``plateau'' of a
SN~IIP that achieves similar peak luminosity.  The energy released by
07qd is substantially lower than that of normal SN~Ia, but the
predicted nickel mass serves as an intermediate example between 08ha
and 05hk, further linking the two.

When compared with other 02cx-like SN, spectra of 07qd mirror that of
08ha and display striking relationships to others.  These comparisons
also imply that strong Si~II is possible in 02cx and 05hk, though
``disguised" by Fe~II blending due to their higher photospheric
velocities.  The fast evolution of 07qd suggests that these IMEs are
only detectable for a brief time, and become masked by Fe~II or
recombine as the photosphere slows and cools.  The presence of carbon
and oxygen ions in the photosphere echoes the results of deflagration
models including those of \cite{gamezo04}, suggesting the presence of
unburned white dwarf material and supporting that this class stems
from such a progenitor.

Also of interest is the ``SN .Ia'' model, which concerns doubly
degenerate white dwarf pairs undergoing a helium flash strong enough
to produce $^{56}$Ni and absolute $V$ magnitudes as low as $-15$ mag
\citep{bildsten07}.  \citet{shen10} used this model to predict the
luminosities and rise times that we observe, but also spectra
dominated by the He-burning products Ca and Ti (detected but not
prominent).  Combined with the low observed expansion velocities, we
find 07qd to be an unlikely SN .Ia candidate.

Our estimated radioactive nickel yield for 07qd ($\sim0.01$
M$_{\odot}$) is extremely small, and it is difficult to understand how
such a weak thermonuclear explosion could disrupt a Chandrasekhar-mass
white dwarf. \citet{woosley07} found that a minimum of $4.6 \times
10^{50}$~ergs of nuclear energy is required to unbind a
Chandrasekhar-mass white dwarf, and that can be achieved by producing
0.3~M$_\odot$ of $^{56}$Ni. This amount is more than ten times the
$^{56}$Ni yield we estimate for 07qd. Of course, nonradioactive
elements might dominate the nuclear energy term; for example, the
production of 0.37~M$_\odot$ of IMEs could also disrupt the star. However,
the synthesized elements in three-dimensional pure-deflagration models
discussed by \citet{blinnikov06} are dominated by the iron peak,
especially $^{56}$Ni, over IMEs.  It is also possible that the synthesis 
of nonradioactive $^{58}$Ni or other stable iron-group elements may
supply at least some of the kinetic energy.

To estimate the nickel yield, we have applied ``Arnett's rule''
\citep{arnett82}, which allows us to use the peak luminosity as an
indicator of the radioactive energy deposition. The rule assumes that
nearly all of the energy produced by nickel and cobalt decays is
trapped and converted to ultraviolet, optical, and infrared radiation.
Calculations by \citet{pinto00a} indicate that complete trapping is an
excellent approximation for the first month after the explosion when
the nickel distribution is concentrated toward the center.

Deflagration models suggest that Rayleigh-Taylor instabilities cause
hot nickel bubbles to rise as cooler unburned material sinks.
\citet{blinnikov06} find fairly uniform distributions of $^{56}$Ni in
radially averaged three-dimensional pure-deflagration
simulations. This leaves more radioactive nickel in the outer layers
of the ejecta than predicted in detonation models and implies that
some of the radioactive energy will be lost.  Additionally,
\citet{calder04} suggest that highly asymmetric deflagrations could
both unbind a white dwarf and provide the mechanism through which
heavy iron-group elements may be transported to the surface without a
detonation.

The ejected mass estimated for 07qd is very low, as is the case for 
08ha \citep{foley09,foley10}, perhaps suggesting a sub-Chandrasekhar 
mass explosion. However, even normal SN~Ia events such as SN~1992A 
\citep{cappellaro97,stritzinger06} appear to have ejected less than a
Chandrasekhar mass of material based on similar assumptions. 91bg-like
events also give significantly low ejected masses from the methods
currently being applied. We suspect that some important physics is
being ignored in the ejected mass calculations and the results
should be considered ambiguous.

The presence of cobalt and high photospheric temperatures in the
early-time spectra support the conjecture that radioactive elements
were mixed up to the outer layers of 07qd.  But \citet{pinto00a} and
\citet{jeffery99} show that a uniform $^{56}$Ni distribution should
only result in a 10\% to 30\% error in the nickel-mass estimate from
Arnett's rule, not the order of magnitude needed to reconcile the
energetics.  An ``inverted'' SN~Ia, where most of the nickel is in the
outer layers of the ejecta (assume \citealt{jeffery99} $q$ values are
$< 1/3$), would result in inefficient $\gamma$-ray deposition and
extreme violation of Arnett's rule.  Such a $^{56}$Ni distribution,
however, is likely to result in the ejecta turning optically thin to
$\gamma$-rays very early.

\begin{figure}[h]
\epsscale{.7}
\plotone{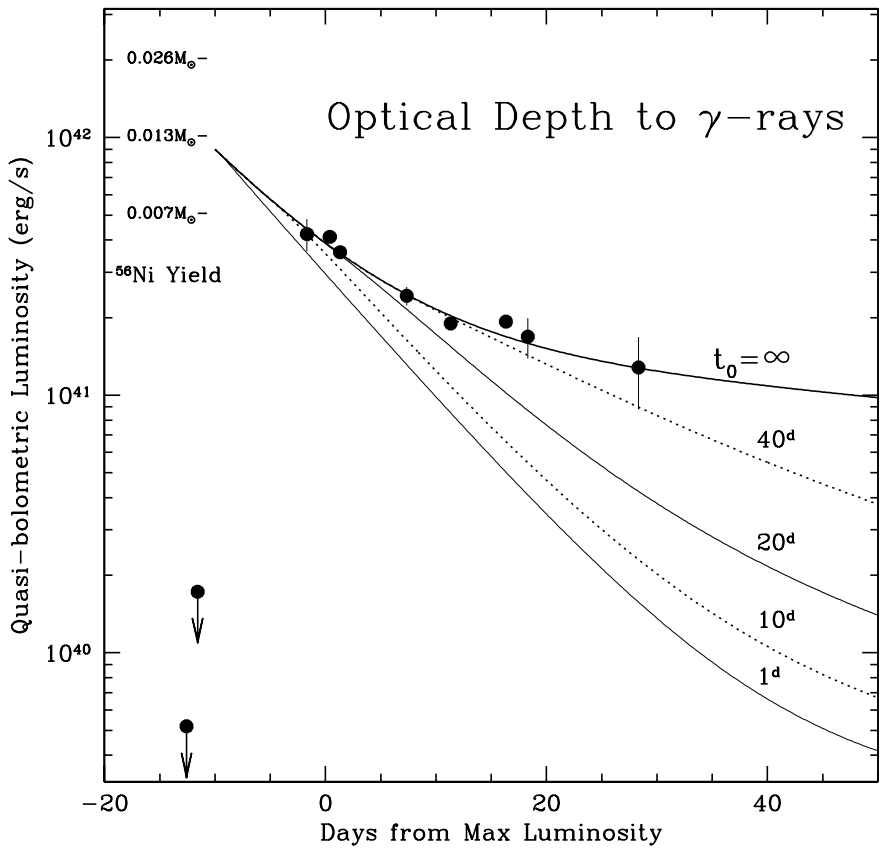}
\caption{\label{f:bol_luminosity} The quasi-bolometric luminosity of
  07qd estimated from the SDSS-II $ugriz$ photometry versus days
  from maximum light. The arrows indicate luminosity upper limits from
  nondetections. The lines show models of radioactive energy
  deposition which vary in their time to become optically thin to
  $\gamma$-rays \citep{jeffery99}. The models fade faster than the
  observations for $t_0 < 40$~days, implying that the $\gamma$-ray
  trapping is efficient for a month after maximum light, similar to
  the fiducial time for normal SN~Ia.}
\end{figure}

\citet{stritzinger06} used the characteristic time for SN~Ia to become
optically thin to $\gamma$-rays, $t_0$, as a way to estimate the
ejected mass and kinetic energy. They found that $t_0$ ranges between
25 and 35 days for normal SN~Ia, coinciding with a range of ejecta and nickel masses.   
\citet{jeffery99} showed that for a
fixed ejecta mass, opacity, and density structure, $t_0 \approx
q^{1/2}$, so pushing radioactive nickel away from the center (smaller
nickel concentration parameter $q$) will shorten the time it takes for
the ejecta to become optically thin. Figure \ref{f:bol_luminosity}
shows the quasi-bolometric luminosity curve (constructed from $ugriz$
photometry) for 07qd and displays radioactive energy deposition curves
parameterized by the fiducial time, $t_0$. For high optical depths,
the curves should obey Arnett's rule and pass through the light-curve
maximum. The models that become optically thin to $\gamma$-rays soon
after explosion have the freedom to be shifted to higher nickel yields
as the radioactive energy deposition will match the luminosity after
maximum light.  The best match to the observations is for $t_0 >
40$~days, suggesting the ejecta are efficiently trapping radioactive
energy a month after maximum brightness.  The optically thin curves
decline much faster than the data, implying that Arnett's rule applies
even for this weak explosion.  02cx and 05hk, measured more than 200
days past maximum, continued to exhibit narrow P-Cygni profiles in
their spectra \citep{jha06,sahu08}.  08ha appears to be consistent
with this \citep{foley10}, so 07qd may be relatively opaque to optical
photons and $\gamma$-rays for some time.

 \subsection{Conclusions}

Analyzing the photometry and spectroscopy of 07qd, we find the
following.

(1) 07qd was spectroscopically similar to both 08ha and 05hk.  Strong
lines of Fe~II and Co~II are present in spectra of all three objects,
while Fe~III and IME features are most visible at early epochs, though
their optical depths decline quickly.

(2) The explicit presence of a variety of IMEs in the early-time
spectra implies the thermonuclear deflagration of carbon and oxygen
and shows 07qd to be inconsistent with a core-collapse event.

(3) Correlations exist between 02cx-like peak luminosity, photospheric
velocity, and light-curve stretch, and the events span a sequence
from 08ha to 02cx.

(4) The diminutive peak luminosity of 07qd implies that the $^{56}$Ni
yield was quite small, lying between that of 08ha and 05hk.

Findings (1) and (2) point toward thermonuclear burning during the
explosion of 07qd, while points (3) and (4) emphasize that 07qd
completes a sequence of 02cx-like SN~Ia both energetically and
spectroscopically.  The composition and velocity of the ejecta support
the picture of a deflagration.  However, deflagration models such as
$1\_3\_3$ used by \citet{blinnikov06} do not predict the short rise
time and small nickel mass seen in 07qd or 08ha.  The small amounts of
synthesized radioactive nickel argue against the unbinding of a
Chandrasekhar-mass progenitor, but if the fusion is dominated by IMEs
at the expense of nickel, the explosions of massive white dwarfs are
still possible.  Other possible scenarios are as follows.

(a) The explosions are thermonuclear with sub-Chandrasekhar masses.
This would decrease the progenitors' gravitational binding energies,
thus lowering the ejecta masses and internal energies needed to
disrupt the stars.

(b) Arnett's rule does not apply to this class of objects, implying that
the $^{56}$Ni mass is significantly underestimated.  The bolometric
light curve of 07qd, however, suggests that Arnett's rule is
operational to some degree.

(c) The explosions are actually core-collapse scenarios that have
somehow managed to synthesize significant amounts of Si and S.

The low velocities and energies present in 08ha and 07qd enable the
analysis of many aspects of 02cx-like SN, otherwise hidden by the
Fe~II blending present in 05hk and 02cx.  Though the velocities and
energies span a wide range, together they constitute a well-defined
group of peculiar SN~Ia.

\acknowledgments

Funding for the SDSS and SDSS-II has been provided by the Alfred
P. Sloan Foundation, the Participating Institutions, the National
Science Foundation (NSF), the U.S. Department of Energy (DOE), the
National Aeronautics and Space Administration (NASA), the Japanese
Monbukagakusho, the Max Planck Society, and the Higher Education
Funding Council for England. The SDSS Web Site is
http://www.sdss.org/.

The SDSS is managed by the Astrophysical Research Consortium for the
Participating Institutions. The Participating Institutions are the
American Museum of Natural History, Astrophysical Institute Potsdam,
University of Basel, University of Cambridge, Case Western Reserve
University, University of Chicago, Drexel University, Fermilab, the
Institute for Advanced Study, the Japan Participation Group, Johns
Hopkins University, the Joint Institute for Nuclear Astrophysics, the
Kavli Institute for Particle Astrophysics and Cosmology, the Korean
Scientist Group, the Chinese Academy of Sciences (LAMOST), Los Alamos
National Laboratory, the Max-Planck-Institute for Astronomy (MPIA),
the Max-Planck-Institute for Astrophysics (MPA), New Mexico State
University, Ohio State University, University of Pittsburgh,
University of Portsmouth, Princeton University, the United States
Naval Observatory, and the University of Washington.

Some of the data presented herein were obtained at the W. M. Keck
Observatory, which is operated as a scientific partnership among the
California Institute of Technology, the University of California, and
NASA; it was made possible by the generous financial support of the
W. M. Keck Foundation. The authors wish to recognize and acknowledge
the very significant cultural role and reverence that the summit of
Mauna Kea has always had within the indigenous Hawaiian community; we
are most fortunate to have the opportunity to conduct observations
from this mountain. We thank the Keck staff for their assistance.

We are grateful for the financial support of the University of Notre
Dame, NASA/STScI Grant HST-GO-10893.01-A to C.M.M., the NSF, and the
DOE, specifically NSF grant AST--0908886 and DOE grant
DE-FG02-08ER41563 to A.V.F.  Support for this research at Rutgers University 
was provided by DOE grant DE-FG02-08ER41562 and NSF award AST--0847157 to 
S.W.J.  We thank Brian Hayden, Joe Gallagher, and
Weidong Li for discussions and their help in the production of this
paper, Jerod Parrent and the Online Supernova Spectrum Archive
(SUSPECT), and David Jeffery along with the SUSPEND database.

\clearpage
\begin{deluxetable}{cccccc}
\tabletypesize{\scriptsize}
\tablecaption{\label{t:phot_data} Observed SDSS Photometry of 07qd in 
Flux Densities\tablenotemark{a,b} }
\tablewidth{0pt}
\tablehead{
\colhead{MJD}	&	\colhead{$u$ [$\mu$Jy]} & \colhead{$g$} & \colhead{$r$} & \colhead{$i$} & \colhead{$z$}  
}
\startdata
54346.41 & -0.010 $\pm$  2.726 &  1.620 $\pm$  1.230 &  0.880 $\pm$  1.235 & -0.730 $\pm$  1.281 & -1.100 $\pm$  3.851 \\ 
54348.41 & -1.780 $\pm$  1.614 &  0.610 $\pm$  0.648 &  1.520 $\pm$  0.698 &  0.200 $\pm$  1.019 &  1.600 $\pm$  3.929 \\ 
54355.42 &  1.500 $\pm$  2.146 & -0.470 $\pm$  0.530 &  0.410 $\pm$  1.111 & -0.370 $\pm$  1.813 &  5.870 $\pm$  5.868 \\ 
54358.37 &  4.310 $\pm$  1.878 &  0.870 $\pm$  0.530 & -0.000 $\pm$  0.707 & -1.480 $\pm$  1.159 & -1.940 $\pm$  4.186 \\ 
54365.40 & 0.000 $\pm$  0.971 & -0.180 $\pm$  0.315 & 0.000 $\pm$  0.496 & -1.500 $\pm$  0.719 & -1.470 $\pm$  3.225 \\ 
54381.42 &  1.400 $\pm$  1.165 &  0.010 $\pm$  0.362 & -0.090 $\pm$  0.590 & -0.340 $\pm$  0.790 & -4.300 $\pm$ 3.010 \\ 
54384.43 &  1.550 $\pm$  1.396 &  0.130 $\pm$  0.393 & -0.580 $\pm$  0.557 & -2.860 $\pm$  0.817 & -5.780 $\pm$ 3.438 \\ 
54386.41 &  0.260 $\pm$  1.478 &  0.080 $\pm$  0.489 &  0.230 $\pm$  0.634 &  0.670 $\pm$  0.846 & -1.960 $\pm$ 4.057 \\ 
54388.42 &  0.240 $\pm$  1.431 &  0.270 $\pm$  0.399 &  0.060 $\pm$  0.587 & -0.250 $\pm$  0.779 &  2.090 $\pm$ 3.177 \\ 
54392.42 & -0.840 $\pm$  1.554 & -0.420 $\pm$  0.556 &  1.160 $\pm$  0.837 &  0.120 $\pm$  1.228 & -1.630 $\pm$ 4.251 \\ 
54393.42 &  2.100 $\pm$  1.731 & -0.020 $\pm$  0.564 & -0.670 $\pm$  0.546 & -0.050 $\pm$  0.850 &  3.650 $\pm$ 4.373 \\ 
54396.29 & -7.950 $\pm$  7.556 & -2.870 $\pm$  3.713 &  0.720 $\pm$  2.329 & - 			& -3.160 $\pm$ 7.688 \\ 
54405.39 & 13.620 $\pm$  2.416 & 19.170 $\pm$  1.007 & 16.510 $\pm$  0.990 & 15.750 $\pm$  1.281 &  7.640 $\pm$ 4.463 \\ 
54406.33 &  6.930 $\pm$  1.466 & 17.310 $\pm$  0.638 & 17.080 $\pm$  0.709 & 15.930 $\pm$  0.912 & 17.020 $\pm$ 2.989 \\ 
54412.35 &  2.470 $\pm$  1.702 &  9.760 $\pm$  0.514 & 15.720 $\pm$  0.899 & 13.850 $\pm$  1.204 & 15.080 $\pm$ 4.166 \\ 
54416.32 &  0.920 $\pm$  1.072 &  6.350 $\pm$  0.388 & 14.140 $\pm$  0.574 & 13.620 $\pm$  0.769 & 17.010 $\pm$ 3.035 \\ 
54421.33 &  3.480 $\pm$  1.736 &  5.250 $\pm$  0.502 & 13.290 $\pm$  0.797 & 11.980 $\pm$  1.110 & 12.360 $\pm$ 4.757 \\ 
54423.31 &  1.060 $\pm$  1.924 &  4.200 $\pm$  0.829 & 12.460 $\pm$  1.058 & 13.620 $\pm$  1.197 & 19.620 $\pm$ 3.800 \\ 
54433.33 & -				&  5.430 $\pm$  1.501 &  8.780 $\pm$  2.494 &  9.740 $\pm$  2.832 &  5.980 $\pm$ 6.150 \\ 
\enddata
\tablenotetext{a}{Not corrected for Milky Way extinction.}
\tablenotetext{b}{Data associated with poor seeing have been omitted from this list.}
\end{deluxetable}

\clearpage
\begin{deluxetable}{c c c c c c}
\tabletypesize{\scriptsize}
\tablecaption{\label{t:spec_sched} Journal of Spectroscopic Observations}
\tablewidth{0pt}
\tablehead{\colhead{Telescope}	&	\colhead{UT Date}	&	\colhead{Time [UT]}	&	\colhead{Days since $B_{\rm max}$}	&	\colhead{Exposure [s]}	&	\colhead{Range [\AA]}
}
\startdata
TNG		&	5 Nov. 2007	&	02:06:10	&	+3	&	3 $\times$ 1800			&	3673--7401\\
HET		&	10 Nov. 2007	&	04:23:12	&	+8	&	1200			&	4075--9586\\
Keck		&	12 Nov. 2007	&	12:39:01	&	+10	&	1500			&	3073--8800\\
HET		&	17 Nov. 2007	&	03:58:29	&	+15	&	1200			&	4074--9586\\
\enddata
\end{deluxetable}

\clearpage
\begin{deluxetable}{r c c c c c}
\tabletypesize{\scriptsize}
\tablecaption{\label{t:qd_3d} SYNOW Parameters for Figure \ref{f:3d_decomposed}\tablenotemark{a,b}}
\tablewidth{0pt}
\tablehead{
\colhead{Ion} & \colhead{$\tau$} & \colhead{$v_{\rm min}$} &  \colhead{$v_{\rm max}$} & \colhead{$v_e$} & \colhead{$T_{\rm exc}$}
}
\startdata
      Na I &      0.3 &       4.5 &       8.0 		&        4 &       10 \\
      Na I &      0.5 &       2.8 &       4.5 		&        1 &       10 \\
      Na I &      0.1 &       8.0 &       10. 		&        2 &       10 \\
      Ca II &       10 &       2.8 &       3.0 	&        2 & 	12 \\
      S II &       2.7 &       2.8 &       3.3 		&        2 & 	10 \\
      Si II &       3.0 &       2.0\tablenotemark{c} &      $\infty$ &        1 & 	8 \\
      Si III &	   2.0 &		2.8 &	   $\infty$ &		 1 &	10 \\
      Fe III &      0.7 &       2.8 &     $\infty$ &        1 & 	10 \\
      Fe II &       1.0 &       2.8 &      $\infty$ &        1 & 	10 \\
      Co II &       3.0 &       2.8 &      $\infty$ &        1 & 	10 \\
       C II &	        0.02 &    2.8 &     $\infty$ &         1 &         10 \\
       C III &      0.5 &       5.0 &       8.0 	&        2 & 	10 \\
       O I &       9.5 &       2.8 &       6.5 		&        3 &       10 \\
       O III &       1.8 &       2.8 &     $\infty$ &        1 & 	10 \\
      Mg I &      0.5 &       2.8 &       $\infty$ &        2 &       12 \\
      Ti I &       1.0 &       2.8 &       $\infty$ &        1 &       10 \\
      Cr I &       2.0 &       2.8 &       4.0 		&        2 &       10 \\
      Ca I &      2.0  &       5.0 &       $\infty$        &        1         &      10 \\
      Al I &	     3.0 &        2.8 &      $\infty$ &        1        &        10 \\
\enddata
\tablenotetext{a}{Photospheric velocity of 2800 km $\mathrm{s}^{-1}$
  and a blackbody temperature of 10,000 K.}
\tablenotetext{b}{Velocities are given in units of 1000 km
  $\mathrm{s}^{-1}$ and $T_{\rm exc}$ values are given in units
  of 1000 K.}
\tablenotetext{c}{Si~II simulated individually at 2000 km s$^{-1}$}
\end{deluxetable}

\clearpage
\begin{deluxetable}{r c c c c c}
\tabletypesize{\scriptsize}
\tablecaption{\label{t:qd_10d} SYNOW Parameters for Figure \ref{f:qd_synow_10d}\tablenotemark{a,b}}
\tablewidth{0pt}
\tablehead{
\colhead{Ion} & \colhead{$\tau$} & \colhead{$v_{\rm min}$} & \colhead{$v_{\rm max}$} & \colhead{$v_e$} & \colhead{$T_{\rm exc}$}
}
\startdata
Co II		&	10		&	2.8				&	3.8				&	1		&	8				\\
Fe II		&	10		&	2.8				&	3.8				&	1		&	8				\\
Ca II		&	35		&	2.8				&	4.8				&	1		&	8				\\
Na I		&	0.4		&	2.8				&	4.8				&	1		&	8				\\
Na I		&	0.3		&	4.0				&	7.0				&	8		&	8				\\
Na I		&	0.3		&	7.0				&	11.0				&	3		&	8				\\
Si II		&	3.0		&	2.8				&	3.8				&	1		&	8				\\
Sc II		&	1.0		&	2.8				&	5.8				&	1		&	10				\\
O I		&	1.0		&	2.8				&	$\infty$			&	1		&	8				\\
O II		&	0.2		&	3.5				&	$\infty$			&	1		&	8				\\
O III		&	2.0		&	2.8				&	$\infty$			&	1		&	8				\\
\enddata
\tablenotetext{a}{Photospheric velocity of 2800 km $\mathrm{s}^{-1}$
  and a blackbody temperature of 8,000 K.}
\tablenotetext{b}{Velocities are given in units of 1000 km
  $\mathrm{s}^{-1}$ and $T_{\rm exc}$ values are given in units of
  1000 K.}
\end{deluxetable}

\end{document}